\documentclass[fleqn,usenatbib,useAMS]{mnras}

\usepackage{graphicx}	
\usepackage{amsmath}	
\usepackage{amssymb}	
\usepackage{multicol}        
\usepackage{bm}		
\usepackage{pdflscape}	

\newcommand{\kms}{${\rm \,km\,s}^{-1}$} 

\usepackage[T1]{fontenc}
\usepackage{ae,aecompl}

\usepackage{newtxtext,newtxmath}

\title[A double degenerate merger in the Milky Way halo]{An ancient double degenerate merger in the Milky Way halo}

\author[A. Kawka et al.]{Adela Kawka$^{1}$\thanks{Contact e-mail: \href{mailto:adela.kawka@curtin.edu.au}{adela.kawka@curtin.edu.au}}, St\'ephane Vennes$^{1, 2}$ and
Lilia Ferrario$^{1, 2}$
\\
$^1$ International Centre for Radio Astronomy Research - Curtin University, GPO Box U1987, Perth, WA 6845, Australia\\
$^2$ Mathematical Sciences Institute, The Australian National University, Canberra, ACT 0200, Australia}

\date{Last updated 2019 August 8; in original form 2019 August 8}

\pubyear{2019}

\begin{document}
\label{firstpage}
\pagerange{\pageref{firstpage}--\pageref{lastpage}}
\maketitle

\begin{abstract}

We present an analysis and re-appraisal of the massive, carbon-enriched (DQ) white dwarf (WD) LP~93-21. Its high mass ($\approx 1~M_\odot$) and membership to the class of warm DQ WDs, combined with its peculiar halo kinematics suggest that this object is the product of an ancient stellar merger event, most likely that of two WDs. Furthermore, the kinematics places this object on a highly retrograde orbit driven by the accretion of a dwarf galaxy onto the Milky Way that occurred at a red shift greater than 1.5. As the product of a stellar merger LP~93-21 is probably representative of the whole class of warm/hot DQ WDs.
\end{abstract}


\section{Introduction}

White dwarfs are the final stage in the life of a majority of stars, and many are found in close binary systems, either in double degenerate binaries or as a companion to a main-sequence star. However, the fraction of binaries in the WD population is only about 25\% which is significantly lower than the fraction of about 50\% of binaries among main-sequence stars \citep{Ferrario2012, too2017} thus indicating that there should be a population of single WDs that formed through merging events described by, e.g., \citet[][]{lor2009,Briggs2015}. 

The presence of an excess of massive WDs in the local population was first uncovered in the Extreme Ultraviolet (EUVE) survey of hot WDs \citep{ven1997,ven1999}. Recently, using Gaia parallax measurements
and Sloan Digital Sky Survey (SDSS) photometric measurements, \citet{kil2018} found that the WD mass distribution is bifurcated and that the massive WD peak near $0.8\,M_\odot$ and the high-mass tail beyond harbour WDs that formed in mergers. However, \citet{ber2019} suggested that some of these objects could have a normal mass ($\sim 0.6\,M_\odot$) but with a mixed H/He atmosphere. Further observations would be required to disentangle the effects of mass and chemical composition.

We expect that WDs that are the products of mergers should have properties that differ from those originating from single star evolution. For instance, mergers are likely to be massive, have short rotation periods, and some could be magnetic. In fact, a significant fraction of WDs possess a magnetic field, although the incidence of magnetism varies among the various spectral classes \citep{kaw2019}. One group of WDs, the rare hot and carbon-rich DQs, have an exceptionally high incidence of magnetism \citep{duf2013}. They are also fast rotators, with periods ranging from a few minutes to a couple of days \citep{dun2015,wil2016}, and have an average mass higher than that of the general WD population. Recently, \citet{cou2019} showed that there is a population of massive, warm ($10\,000 \lesssim T_{\rm eff} \lesssim 16\,000$ K) DQs and suggested that these may be the descendants of hot DQs.

The case of the warm DQ LP~93-21 merits further attention. This object was first identified as a high proper motion star in the Luyten Palomar survey and as a WD candidate by \citet{luy1968}. \citet{san1968} first discovered that LP~93-21 could have a retrograde orbit. The first spectra were obtained by \citet{lie1977} and \citet{gre1977}. The latter also found that LP~93-21 is likely to have a retrograde orbit and is more massive than the average WD. A trigonometric parallax of $\pi = 0.0114\pm0.0054$ arcsec was obtained by \citet{har1985}, but the uncertainty was too large to place a useful constraint on its mass. The first modelling of its spectrum was performed by \citet{bue1979}, who measured $T_{\rm eff} = 8500$~K, $\log{g}=7.8$ and $\log{\rm C/He}=-2.9$, resulting in a relatively low mass ($\approx0.5\,M_\odot$). 
However, the recent Gaia parallax measurement firmly established its very high mass \citep[$\approx1.1\,M_\odot$,][]{leg2018,kil2019}.

Recently, LP~93-21 was reported as a hyper-runaway WD and Type Iax supernova (SN Iax) remnant candidate by \citet{ruf2019}. They proposed that LP~93-21 is unbound to the Milky Way and that its large velocity and carbon-rich surface occurred in the aftermath of a partial core deflagration, similar to events leading to the formation of the hyper-velocity and low mass WD LP~40-365 \citep{ven2017}. However, unlike LP~40-365 which has a very peculiar oxygen/neon-rich atmosphere devoid of traces of hydrogen or helium, LP~93-21 is helium dominated with a large quantity of carbon mixed in the convective envelope \citep{leg2018,kil2019} but without the expected products of carbon burning.

In this paper we  show that the stellar properties and kinematics of LP~93-21 are more consistent with this star being the product of a stellar merger. We revisited SDSS spectroscopy of known DQ WDs (Section \ref{obs}) and extracted reliable stellar parameters and radial velocity measurements (Sections~\ref{param} and \ref{carbon_spectrum}). In particular, we pay close attention to the density and temperature in the line forming region and their effect on the carbon line spectrum. We discuss the revised kinematics (Section~\ref{kinematics}) in the context of an ancient dwarf galaxy merger with the Milky Way in Section~\ref{population}. Finally, we identify likely progenitors for the warm DQ WD LP~93-21 in Section~\ref{progenitors} and we summarize our results and conclude in Section~\ref{conclusions}.

\section{Observation and Analysis}\label{obs}

We have assembled the known population of DQ WDs and extracted astrometric and photometric measurements from the second data release of Gaia \citep{gai2018}. We also collected spectroscopic and photometric measurements of LP~93-21 and related stars from the SDSS \citep{agu2019}. Table~\ref{tab_obs} lists astrometric and photometric data for LP~93-21. 

\begin{table}
\centering
\caption{Astrometry and photometry}
\label{tab_obs}
\begin{tabular}{lcc}
\hline
Parameters & Measurement & Reference \\
\hline
RA (J2000)    & 10 45 59.42 & 1 \\
Dec (J2000)   & +59 04 51.32 & 1 \\
$\mu_\alpha \cos{\delta}$ (\arcsec yr$^{-1}$) & $-1.0192\pm0.0001$ & 1 \\
$\mu_\delta$ (\arcsec yr$^{-1}$)              & $-1.4625\pm0.0002$ & 1 \\
$\pi$ (mas)     & $17.482\pm0.135$   & 1 \\
$G$ & $17.677\pm0.002$ & 1 \\
$G_{\rm BP}$ & $17.683\pm0.010$ & 1 \\
$G_{\rm RP}$ & $17.448\pm0.016$ & 1 \\
$u$ & $17.784\pm0.011$ & 2 \\
$g$ & $17.763\pm0.005$ & 2 \\
$r$ & $17.691\pm0.006$ & 2 \\
$i$ & $17.725\pm0.007$ & 2 \\
$z$ & $17.884\pm0.018$ & 2 \\
\hline
\end{tabular}\\
References: (1) \citet{gai2018}; (2) \citet{agu2019}
\end{table}

\subsection{Stellar parameters: LP~93-21}\label{param}

\citet{kil2019} and \citet{leg2018} determined effective temperatures of $8690\pm120$~K and $9730\pm49$~K, respectively and high masses of $1.03\pm0.02$~$M_\odot$ and $1.14\pm0.01$~$M_\odot$, respectively. We have re-analyzed the SDSS spectrum and Gaia parallax of LP~93-21 using models described in \citet{ven2012} and obtained stellar parameters entirely consistent with the above results. Details of the models will be described in a forthcoming paper (Kawka et al. 2019, in preparation). Table~\ref{tab_properties} lists the stellar properties of LP~93-21 including the Gaia distance measurement.

\subsection{The carbon line spectrum}\label{carbon_spectrum}

We examined the SDSS spectra of a sample of nine warm, massive DQ WDs (10-16$\times10^3$~K, $M>0.8M_\odot$) including LP~93-21. These objects have a higher carbon abundance ($\log{\rm C/He}\approx -4$ to $-2$) and a higher mass than normal DQ WDs \citep{cou2019}, but they are somewhat cooler than the hot DQ WDs described by \citet{duf2008}. The optical spectra show neutral carbon lines and, in cooler objects, strong molecular carbon bands. The SDSS spectrum of LP~93-21 shows strong C$_2$ Swan bands and \ion{C}{i} lines at $\lambda_{\rm vac}=4933.40, 5381.83$\AA\ together with prominent multiplets at {\it 4770.70}, {\it 7116.44} and {\it 7117.17}\AA. Additional strong lines detected in the SDSS spectra of other warm DQ WDs include \ion{C}{i} $\lambda_{\rm vac} = 4270.22$ and 8337.44\AA\ and the multiplet at {\it 9089.34}\AA.

Radial velocity measurements in warm DQ WDs show large shifts between carbon band heads and atomic lines and between groups of atomic lines due to the Stark effect. Earlier applications of Stark shifts to heavy element line spectra include the analysis of the DAZ WD GALEX~J1931+0117 \citep{ven2011}. The evidence for Stark shifts in the \ion{C}{i} lines is exposed in the following comparison between laboratory and stellar measurements. We found that in warm DQ WDs carbon lines with highly excited upper levels, i.e., with an ionization energy $E_{\rm ion}<1.2$~eV, are systematically red shifted by $\approx100$\kms\ relative to lines with lower-lying upper levels, i.e., $E_{\rm ion}>1.2$~eV. In the high electronic density ($n_e\approx10^{17}$~cm$^{-3}$) environment of these DQ WDs the carbon lines are shaped by the Stark effect. Shock tube experiments \citep{mil1970} revealed red shifts of 0.3, 2.2 and 3.0 ($n_e/10^{17}$)~\AA\,${\rm cm}^{-3}$ for three lines of interest at $\lambda_{\rm vac} = 5381.83, 5053.56$ and 4933.40\AA. These measurements were corroborated and supplemented by \citet{mij1995a,mij1995b} and a critical review of \ion{C}{i} Stark shifts was presented by \citet{kon2002}. Applied to the sample of nine objects, the observed velocity shift between the \ion{C}{i} lines at $\lambda_{\rm vac}=$ 4933.40 and 5381.33\AA\ is $\Delta\varv=106$~\kms\ and corresponds to an average electronic density in the line forming region of $0.7\times10^{17}~{\rm cm}^{-3}$. Our model atmosphere calculations show that optical depth unity in the atmosphere of these objects, and of LP~93-21 in particular, is achieved in an environment with a neutral helium density of $n({\rm He})=3-4\times10^{20}~{\rm cm}^{-3}$ and an electronic density $n_{\rm e} = 0.6-1.3\times10^{17}~{\rm cm}^{-3}$ in overall agreement with densities extracted from the measured average Stark shift.

We used the SDSS spectrum of LP~93-21 to measure the WD's radial velocity. Following the procedure described above, we first measured the relative shift between \ion{C}{i} $\lambda_{\rm vac} = 5381.83$ and 4933.40\AA, $\Delta\,\varv = 118$~\kms, from which we derived an electronic density $n_{\rm e}=0.84\times10^{17}~{\rm cm}^{-3}$ and, therefore, a Stark shift $d(\lambda5381)=15.6$~\kms. Subtracting the Stark shift from the raw velocity of $134\pm13$~\kms, we obtain an apparent velocity $\gamma=118\pm16$~\kms. Finally, subtracting a gravitational red shift $\gamma_g=103\pm14$~\kms, obtained by averaging mass estimates from \citet{leg2018}, \citet{kil2019} and the present work, we find the corrected radial velocity $\varv_r = \gamma-\gamma_g=15\pm21$. Interestingly, the observed shift in the C$_2$ bands corrected for the gravitational red shift is $\gamma=-11\pm21$~\kms\ revealing a small pressure shift toward the blue $d({\rm C}_2)\approx -26$~\kms. 
  
In summary, we find that LP~93-21 has a large tangential velocity $\varv_{\rm tan} = 4.74 \mu/\pi = 483$~\kms, but a much smaller radial velocity $\varv_r = 17$~\kms\ than assumed by \citet{ruf2019}. Unfortunately, the rational for their assumed velocity ($472\pm20$~\kms) is obscure since neither cited sources explicitly provide radial velocity measurements and their methodology is unspecified. We can safely assume that neither pressure shift nor gravitational redshift were taken into account by \citet{ruf2019}.

The radial velocities for the other members of the sample and for members of the population of hot DQ WDs were calculated in a similar fashion (Kawka et al. 2019, in preparation). 

\subsection{Kinematics}\label{kinematics}

We converted radial velocity and proper motion measurements into the Galactic velocity components using \citet{joh1987}. We assumed that the solar motion relative to the local standard of rest is ($U_\odot,V_\odot,W_\odot$) = (11.1, 12.2, 7.3) \kms \citep{sch2010}. The resulting velocity vector $(U,V,W)=(-201,-420,57)$\,\kms\ (Table~\ref{tab_properties}) immediately suggests that LP~93-21 is bound to the Milky Way but also that it does not belong to the Galactic disc but, instead, to the Galactic halo. \citet{sio1988} investigated the kinematical properties of various subgroups of the WD population and noted that DQ WDs have higher than average space velocities and a higher fraction of halo candidates than the other WD subgroups.

Using the velocity components $(U,V,W)$, we calculated the Galactic orbit and kinematic properties using the Numerical Integrator of Galactic Orbits (NIGO) developed by \citet{ros2015} and the Galactic potential described as model A in \citet{con2019}. In this model the total mass of the Milky Way is $9\times10^{11}\,M_\odot$. Fig.~\ref{fig_kin1} shows the calculated $z$-component of the angular momentum, $L_z$, against the orbital eccentricity, $e$, for a sample of 36 hot and warm DQ WDs including LP~93-21. 

\section{Discussion}\label{discuss}

Based on the stellar and kinematical properties of LP~93-21, we now determine its population membership and likely progenitors.

\subsection{Population Membership}\label{population}

\begin{figure}
    \centering
    \includegraphics[viewport=20 160 550 675,clip,width=0.9\columnwidth]{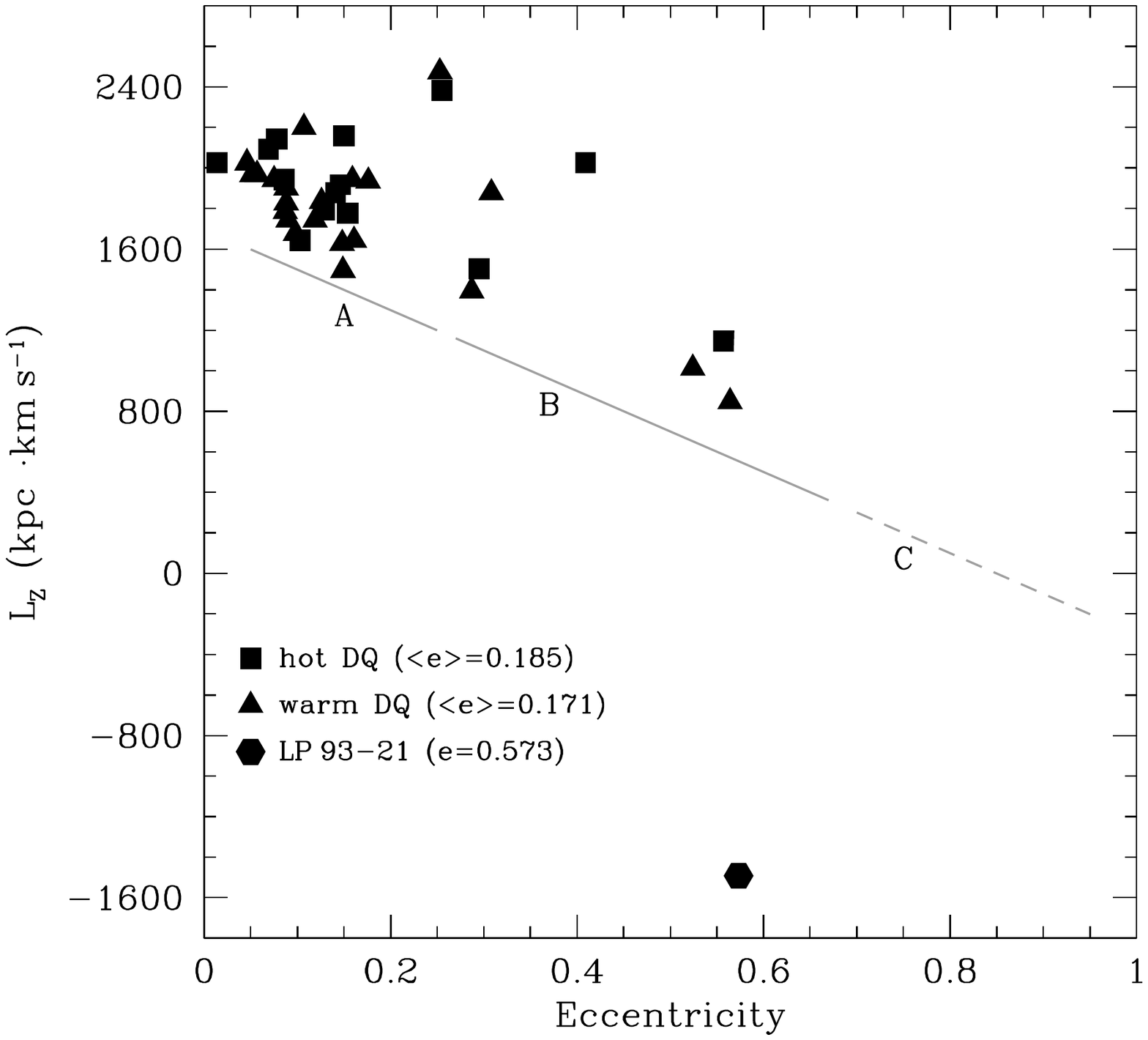}
    \caption{Plot of $L_z$ versus $e$ of hot and warm DQs including LP~93-21. The calibrated locations of thin disc, thick disc, and halo are marked with ``A'', ``B'', and ``C'', respectively, following \citet{pau2006}. The Galactic orbit of LP~93-21 is clearly retrograde.}
    \label{fig_kin1}
\end{figure}

\citet{pau2003} presented a method to clearly distinguish halo WDs from WDs belonging to the thin and thick discs using $L_z$ and $e$. Fig.~\ref{fig_kin1} shows that a few objects in the warm/hot DQ population belong to the thick disc and that LP~93-21 exhibits unique halo properties with a retrograde orbit.  \citet{pau2006} showed that 2 per cent of WDs belong to the Galactic halo including a few objects with retrograde orbits like that of LP~93-21.

\begin{figure}
    \centering
    \includegraphics[viewport=20 400 550 680,clip,width=0.9\columnwidth]{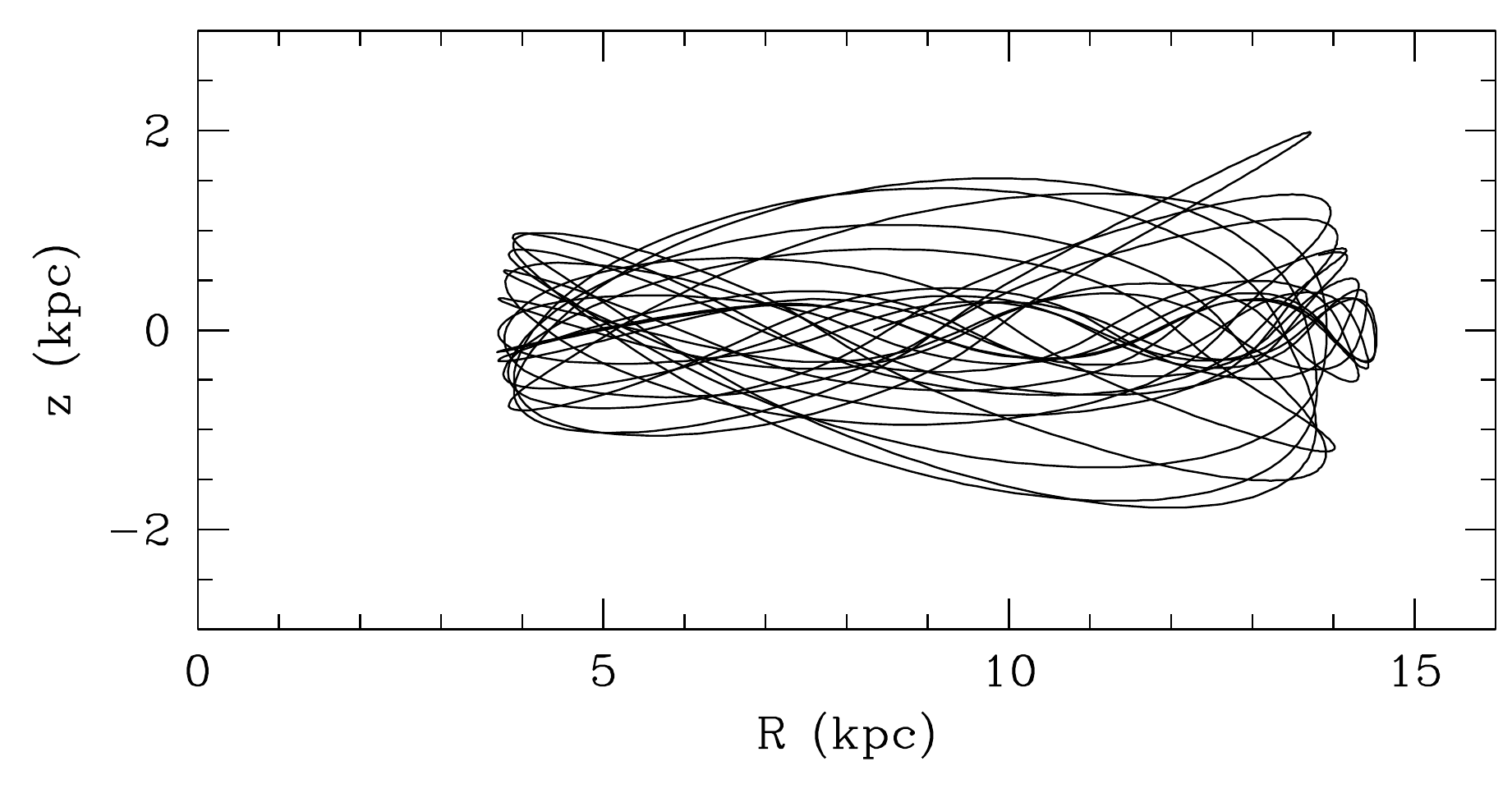}
    \caption{Orbits of LP~93-21 ($0<t<3$ Gyr) in the meridional plane: Galactic height (z) versus radial distance $R=\sqrt{x^2+y^2}$}.
    \label{fig_kin2}
\end{figure}

Fig.~\ref{fig_kin2} shows the calculated orbit of LP~93-21 in the meridional $R$-$z$ plane and Table~\ref{tab_kine} summarizes its orbital properties.
The orbit appears to have some of the attributes of an inner-halo object although its highly retrograde motion points to a different origin for LP~93-21. The results of \citet{matsuno2019} are important for this work because they allow us to understand to which population LP~93-21 may belong. \citet{matsuno2019} divide the $E$-$L_z$ plane in four regions, namely, (A) the innermost halo, whose objects are characterized by small energies $E$ and a prograde motion, (B) the Gaia Enceladus \citep{Helmi2018}, with high $E$ and low $L-z$ objects, (C) the high $E$ retrograde motion stars and (D) the high $E$ and prograde motion stars (only selected to conduct comparison studies to region (C) and thus of no interest in the present context).
The energy of LP~93-21 ($E=-1.4\times 10^5$\,km$^2$/s$^2$) and its retrograde orbit ($L_z=-1491$\,kpc\kms) place it in the vicinity of \citet{matsuno2019} subgroup (C). Adopting their dispersion values for ($E,L_z$) we find that LP~91-23 lies within $\approx2.7\sigma$ of the average ($E,L_z$), i.e., where 7 out of 1000 members should reside. \citet{matsuno2019} find that population (C) has the lowest metallicity indicating that the genesis of this retrograde population is neither linked to the innermost halo nor to the Gaia Enceladus population. Furthermore, the very low $\alpha$ element abundances of population (C), suggestive of inefficient star formation, indicates that this retrograde population originally belonged to a satellite dwarf galaxy of the Milky Way that was much smaller than the satellite galaxy that produced the Gaia Enceladus. Additionally, the EAGLE simulations of galaxy formation conducted by \citet{Mackereth2019} have shown that stars originating from the earliest accretion events have a median $e$ of about 0.5 (LP~93-21 has $e=0.6$), whilst the median $e$ of the most recently (i.e., at redshifts $z<1.5$) accreted satellites is around 0.8 (but up to $>0.9$). 
Alternatively, LP~93-21 could have formed in the Milky Way but, following a merger with a satellite galaxy, its motion migrated to a retrograde halo orbit \citep{bon2017}. Using simulations, \citet{bon2017} showed that more than 95 per cent of halo stars would have formed before the last significant merger that occurred about 7 Gyr ago. These observational and theoretical results impose a total age larger than 7\,Gyr for LP~93-21.

\begin{table}
\centering
\caption{Properties of LP~93-21.}
\label{tab_properties}
\begin{tabular}{lcc}
\hline
Parameter & Measurement & Reference \\
\hline
$T_{\rm eff}$ (K) & 9730$\pm$49, 8690$\pm$120, 9360$\pm$200 & 1,2,3 \\
$\log{g}$ (cgs)   & 8.90$\pm$0.02, 8.70$\pm$0.02, 8.84$\pm$0.02 & 1,2,3 \\
$\log {\rm C/He}$ & $-2.73$, $-3.51$, $-3.12\pm0.13$ & 1,2,3 \\
Mass ($M_\odot$)  & 1.139$\pm$0.011, 1.029$\pm$0.015, 1.10$\pm$0.01 & 1,2,3 \\
Cooling Age (Gyr) & 2.366$\pm$0.020, 2.71$\pm$0.08, 2.28$\pm$0.10 & 1,2,3 \\
Distance (pc)     & $57.2\pm0.4$ & 4 \\
$\gamma_{\rm g}$ (\kms) & 102.5$\pm$14.2 & 3 \\
$\varv_{\rm r}$ (\kms) & $15\pm21$ & 3 \\
$U,V,W$ (\kms)    & $-201\pm11$, $-420\pm7$, $57\pm16$ & 3\\
  \hline
 \end{tabular}\\
References: (1) \citet{leg2018}; (2) \citet{kil2019}; (3) This work; (4) \citet{gai2018}.
\end{table}

\begin{table}
\centering
\caption{Galactic orbits}
\label{tab_kine}
\begin{tabular}{ccc}
\hline
Parameters & Measurement & Notes \\
\hline
$e$         & $0.573^{+0.058}_{-0.042}$   & averaged   \\
$R_{\mbox{max}}$ (kpc) & $14.53^{+1.12}_{-0.43}$ & averaged   \\
$Z_{\mbox{max}}$ (kpc)& 2.0 & extremum \\
$L_z$ (kpc\kms)& $-1491^{+59}_{-58}$ &   \\
$E$ (${\rm km}^2\,{\rm s}^{-2}$)& $-1.40\pm0.03\times10^5$ & \\
\hline
\end{tabular}\\
\end{table}

\subsection{The progenitor stars of LP~93-21}\label{progenitors}

The progenitor of the DQ WD LP~93-21 cannot be a single main sequence star because its 1.1\,M$_\odot$ mass implies a progenitor of about 6\,M$_\odot$ \citep[e.g.,][]{Ferrario2005} that would evolve to the WD stage in $\lesssim$100\,Myr. Therefore, under the single star evolution scenario, the total age of LP~93-21 could not be significantly larger than its $\sim2.3$\,Gyr cooling age. Furthermore, such a relatively massive star would have been born in the spiral arms of the Galaxy where star formation is thriving. As a consequence, the motion of LP~93-21 should be well and truly confined to the thin disk. The only possible explanation of this age paradox is that LP~93-21 is the end product of binary evolution that terminated with the merging of its two stellar components. The likely progenitor system of LP~93-21 probably consisted of two Carbon Oxygen (CO) WDs that formed many billion years ago and then slowly approached each other under the influence of gravitational radiation to finally merge 2.3\,Gyr ago. We favour this double degenerate merging route because it can easily explain both the large mass of LP~93-21 and its carbon enriched atmosphere. Although the stellar merging hypothesis can solve the single progenitor puzzle, it still cannot explain the highly retrograde motion of LP~93-21. There are two possible explanations of this second puzzle;  either the progenitor stars of LP~93-21 belonged to a satellite galaxy that was accreted by the Milky Way in ancient times or LP~93-21 was born in situ but was later driven onto a highly retrograde halo orbit due to the accretion, that occurred more than 7\,Gyr ago, of a satellite galaxy (see section \ref{kinematics}). Both hypotheses require the age of LP~93-21 to be at least three times larger than implied from single star evolution.

In order to understand the genesis of LP~93-21 we have employed the rapid binary star evolution algorithm, {\sc bse}, of \citet{Hurley2002} and evolved binaries from the main sequence to an age of 11\,Gyr. This age, added to the $\sim 2.3$\,Gyr cooling age, is still below the age of the Galaxy and of its ancient satellites.

When stars evolve off the main sequence, their envelopes expand significantly. If a companion star is present and is on a sufficiently close orbit, the newly formed giant fills its Roche lobe and starts transferring mass to its companion, but at a rate that is far too high for the companion to accept. This results into an ever expanding giant whose outer layers engulf both stars till their envelopes merge into one another. This phase of evolution has been named ``common envelope'' \citep[see][]{Izzard2012}. The {\sc bse} code uses the $\alpha_{\rm CE}$ formalism, where $\alpha_{\rm CE}$ is the common envelope efficiency parameter \citep{Hurley2002}. In the present calculations we have adopted $\alpha_{\rm CE}=0.2$, which is consistent with the values used by \citet{Briggs2015} and \citet{Briggs2018} to model the characteristics of magnetic WDs under the assumption that they originate either via common envelope merging events or double degenerate mergers \citep[see also][]{WTF2014}.

Stellar merging events are common in the Galaxy, as highlighted by the large shortfall of WD binaries in the solar neighbourhood, as first noted by \citet{Ferrario2012} and later modelled by \citet{too2017} in terms of binaries lost to merging events during the course of their evolution. It is unlikely that the incidence of binarity was lower in ancient environments. On the contrary, the detailed studies of \citet{Machida2009} have revealed that given the same median rotation parameter as in the solar neighbourhood, if a gas cloud has a metallicity $<10^{-4}$ most stars are born as members of binary/multiple systems.

Our synthetic population was generated using a Monte Carlo approach. The masses $M_1$ ($\in 1-8$\,M$_\odot$) of the main sequence primaries were drawn from a Salpeter's initial mass function \citep{Salpeter1955} whilst the masses $M_2$ ($\in 0.8-8$\,M$_\odot$) of the companions were drawn from a uniform mass distribution \citep[e.g.][]{Ferrario2012}. The initial orbital periods $P_0$ were drawn from a logarithmically uniform distribution ($2\le \log_{10}(P_0/d)\le 4$). The evolved population was then searched for systems that produced double WDs that merged after a total life-span of around 7 to 11\,Gyr to yield a single massive WD like LP~93-21.  We have found that a typical binary that could be a suitable progenitor of LP~93-21 was initially composed of two stars with masses of 2.1\,M$_\odot$ and 1.8\,M$_\odot$ and an initial period of 1184\,days. After about 1 Gyr, the primary star evolves off the main sequence, becomes a red giant, ignites helium in its core and then evolves to become a super-giant star that loses about 0.3\,M$_\odot$ in stellar winds. During the AGB evolution the stellar envelope expands and forms a common envelope with its companion. This brings the two stars closer together due to friction within the common envelope. The envelope is then ejected exposing the newly formed CO WD. Now the two stars are on a much tighter orbit. The primary is a CO WD with a mass of 0.58\,M$_\odot$, while the secondary, largely unaffected by the events unfolded so far, is still on the main sequence. About 184\,Myr later, the secondary star becomes a red giant, ignites helium in its core and then starts its ascent along the AGB. It is during the AGB evolution that the secondary star initiates a second common envelope at the end of which it will emerge as CO WD with a mass of 0.53\,M$_\odot$. The orbital period is now only 0.344\,days. It has taken the two stars about 1\,696\,Myr to become a double degenerate system. From this point onward the orbit slowly shrinks due to gravitational radiation until, after 9.4\,Gyr, the two CO WDs coalesce to produce a 1.1\,M$_\odot$ WD. After cooling for about 2.3\,Gyr this massive WD with a carbon enriched atmosphere will reach $T_{\mbox{eff}}\sim 10,000$\,K and a total age of 13.3\,Gyr. Although this is probably at the upper end of its possible age, a different combination of initial masses and orbital periods can yield younger objects.

Gaia photometry and parallax measurements have  revealed a pile-up of high-mass WDs \citep{gai2018a} on the Hertzsprung--Russell diagram which has been labelled the Q branch. This has been interpreted by \citet{tre2019} as evidence of WD's crystallization whose onset delays cooling and causes this pile-up.
However, \citet{che2019} have shown that the Q branch stars, of which half are DQs, have significantly older kinematics and therefore require an additional cooling delay of 8 Gyr to explain the pile-up which is most likely caused by the settling of $^{22}$Ne. Because this settling which affects only about 6\% of high mass white dwarfs, favours CO cores and massive CO WDs cannot evolve from single star evolution, \citet{che2019} concluded that those WDs that experience this cooling delay must be the products of WD mergers. According to this scenario, since LP~93-21 is cooler than the WDs on the Q branch, it would have undergone this additional cooling and thus its total cooling age should be $\sim 10$ Gyr rather than $\sim 2.3$ Gyr. If this is correct the stars forming the progenitor binary of LP~93-21 must have been more massive and evolved to the ultimate merger event in less than a few Gyrs to yield a total age that is less than the age of the Galaxy and its ancient satellites. 

\section{Conclusions}\label{conclusions}

We found that the DQ WD LP~93-21 is bound to the Milky Way and that its orbit is quite energetic and highly retrograde. Our kinematical study is based on a careful examination of its carbon line spectrum and a proper evaluation of its radial velocity. The kinematics of LP~93-21 resembles that of a stellar population associated to the ancient merger of a dwarf galaxy with the Milky Way.
The kinematics of LP~93-21 and the inferred antiquity are incompatible with single star evolution which would impose a total age $\lesssim 3$~Gyr. We conclude that this star merged out of the evolution of two CO WDs resulting in a total age $>$10\,Gyr. Our results exclude earlier suggestions \citep{ruf2019} 
that LP~93-21 belongs to a class of surviving remnants of subluminous Type Iax supernova events
\citep{ven2017,rad2019}.

The stellar properties of LP~93-21 do not distinguish it from any other members of the warm/hot DQ population. Its high mass and large carbon abundance are entirely consistent with population average. As the prototype of an old WD population, our study of LP~93-21 suggests that as a whole, warm DQ WDs are the product of stellar mergers, most likely double degenerate mergers.

\section*{Acknowledgements}

LF (2019 ICRAR Visiting Fellow for Senior Women in Astronomy) and SV would like to express their gratitude for the support and hospitality of the staff at the International Centre for Radio Astronomy Research. We would like to thank the staff at Perth Observatory for sharing their library resources. We thank D.T. Wickramasinghe for useful discussions. Funding for the SDSS IV has been provided by the Alfred P. Sloan Foundation, the U.S. Department of Energy Office of Science, and the Participating Institutions. SDSS-IV acknowledges support and resources from the Center for High-Performance Computing at
the University of Utah. The SDSS web site is www.sdss.org. This work presents results from the European Space Agency (ESA) space mission Gaia. Gaia data are being processed by the Gaia Data Processing and Analysis Consortium (DPAC). Funding for the DPAC is provided by national institutions, in particular the institutions participating in the Gaia MultiLateral Agreement (MLA). The Gaia mission website is https://www.cosmos.esa.int/gaia. The Gaia archive website is https://archives.esac.esa.int/gaia.

\bsp	
\label{lastpage}
\end{document}